\pdfoutput=1
\documentclass[letterpaper, 10pt, conference]{ieeeconf}

\usepackage{amsmath,amssymb,graphicx,booktabs,url}

\usepackage{amsthm}

\IEEEoverridecommandlockouts
\newtheorem{theorem}{Theorem}
\newtheorem{lemma}{Lemma}
\newtheorem{corollary}{Corollary}
\newtheorem{proposition}{Proposition}
\theoremstyle{definition}
\newtheorem{definition}{Definition}
\newtheorem{assumption}{Assumption}
\newtheorem{problem}{Problem}
\newtheorem{remark}{Remark}

\newcommand{\IEEEkeywordsblock}[1]{%
  \vspace{2pt}\noindent\textit{Index Terms:} #1\vspace{4pt}}
\newcommand{\E}{\mathbb{E}}
\newcommand{\sgn}{\operatorname{sign}}
\usepackage[hidelinks]{hyperref}
\hypersetup{
  pdftitle={Conformal Koopman Certificates for Nonlinear Multi-Agent Consensus on Time-Evolving Graphs},
  pdfauthor={Indrakshi Dey, Nicola Marchetti},
  pdfkeywords={Multi-agent systems, Koopman operator, conformal prediction, switched systems, data-driven control, consensus}
}

\title{\LARGE\bf
Conformal Koopman Certificates for Nonlinear Multi-Agent Consensus on Time-Evolving Graphs}

\author{Indrakshi~Dey,~\IEEEmembership{Senior Member,~IEEE},
and Nicola~Marchetti,~\IEEEmembership{Senior Member,~IEEE}%
\thanks{This work has been submitted to the IEEE for possible publication.
Copyright may be transferred without notice, after which this version may
no longer be accessible.}%
\thanks{I.~Dey is with the Department of Computing, South East Technological
University, Waterford, Ireland (e-mail: indrakshi.dey@setu.ie).
N.~Marchetti is with the Department of Electrical and Electronic Engineering,
Trinity College Dublin, Ireland (e-mail: nicola.marchetti@tcd.ie).}}
\date{}

\begin{document}
\maketitle

\begin{abstract}
Robot fleets must often agree on a shared value before they act, over
wireless networks that change as the robots move. Their supervisor needs a
deadline: the number of communication rounds after which the fleet has
agreed with a stated probability. Today it comes from simulation plus a
safety margin, with an open probability of being met. We certify it. Our method, conformal Koopman certification, rests on
one observation: averaged over the random network changes, the expected
behaviour of the fleet evolves linearly, for any nonlinearity of the agents.
The decay rate of this linear evolution gives the shape of the deadline. One
remaining constant is calibrated from 200 short test runs by conformal
prediction, a statistical method that turns any predictor into a guarantee
that holds with a user-chosen probability. The certified deadline holds for
every agreement tolerance at once and for any error in the estimated rate.
The same linear model predicts the value the fleet will agree on and
certifies how much corruption of the exchanged messages the collective
decision absorbs. For 60 simulated robots the certified deadline was met in
all 500 test runs against a 90\% target and exceeded the slowest run by
16\%. The classical worst-case bound exceeds that run by more than 160 orders
of magnitude, a linearised deadline is missed in up to 26\% of runs, and direct
conformal calibration costs 80 times more simulation per tolerance.
\end{abstract}

\IEEEkeywordsblock{Multi-agent systems, Koopman operator, conformal
prediction, switched systems, data-driven control, consensus.}

\section{Introduction}
Fleets of tens to hundreds of robots are moving from laboratories to
warehouses, farms and search missions. Before such a fleet acts, its robots
often have to agree on a shared value, for example on how to split a search
area, by exchanging messages over a wireless network that changes as they
move. The supervisor needs a \emph{deadline}: the number of communication
rounds after which the fleet has agreed with a stated probability. A deadline
below the true convergence time makes the fleet act on a split opinion; a
deadline far above it wastes mission time. In current practice the deadline
comes from simulation runs plus a safety margin, and the probability that the
fleet meets it remains open.

\begin{figure*}[t]
\centering
\includegraphics[width=\textwidth]{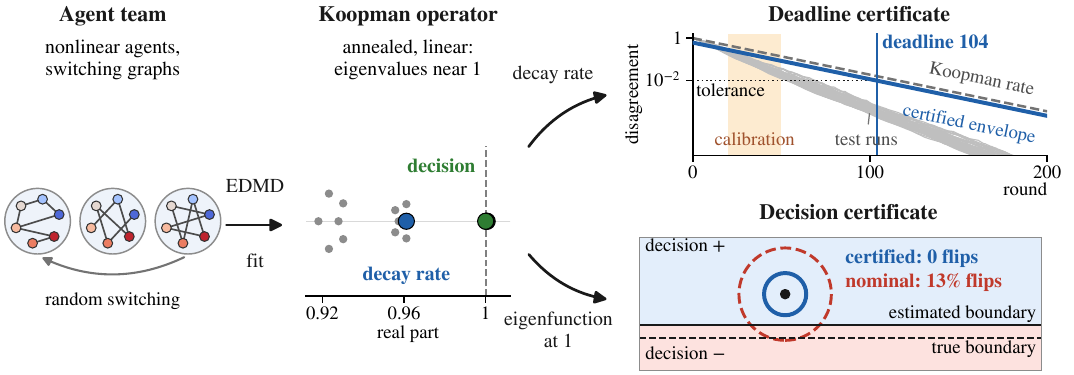}
\caption{Conformal Koopman certification on network~1. EDMD (extended
dynamic mode decomposition) fits the annealed Koopman operator, which gives
the decay rate and the decision eigenfunction; conformal calibration gives
the remaining constant. Decision panel: schematic, flip rates from
Section~\ref{sec:numdec}.}
\label{fig:concept}
\end{figure*}

Two tools have matured that make such a certificate
possible. Koopman operator methods represent a nonlinear system by a linear
operator acting on functions of its state \cite{koopman1931,mezic2005}, and
they now come with data-driven estimators that converge
\cite{williams2015,korda2018} and spectra that can be verified
\cite{colbrook2024}. Conformal prediction (CP) turns any predictor into a
guarantee that holds with a user-chosen probability, from finite data and for
any data distribution \cite{vovk2005,angelopoulos2021}, and it has recently
entered control for certifying stability and safety
\cite{hsu2025,lindemann2024,chee2023}.

A deadline of the form \emph{disagreement decays geometrically} needs a decay rate and the constant in front of it. Classical spectral theory gives the rate for linear teams on
a fixed network. A real fleet is nonlinear, because actuators saturate, and
its network switches at random, so its rate must come from a linear
description of a nonlinear random system. The constant is open even for linear teams: the available bound, the
condition number of an eigenvector matrix, inflates the deadline $1.9$ to
$3.2$ times in our tests (Section~III). Each tool alone covers one half of
the problem: Koopman methods give the shape of the decay and leave its
constant open, and CP gives a constant with a guarantee and needs a shape to
extrapolate across tolerances. We state three requirements, made formal in
Problem~\ref{prob}. (R1) \emph{Coverage}: the fleet meets the deadline with
probability at least $1-\delta$, for a miscoverage level $\delta\in(0,1)$
chosen by the user. (R2) \emph{Length}: the deadline stays close to the
largest convergence time. (R3) \emph{Reuse}: one calibration serves every
agreement tolerance. Existing methods meet two of the three.
\emph{Worst-case ergodicity bounds}
\cite{jadbabaie2003,olfati2004,ren2005,nedic2015} meet R1 and R3, with
deadlines above $10^{164}$ rounds for $60$ agents. \emph{Spectral analysis of
a linearised model} meets R2 and R3, and its deadline is missed in up to
$26\%$ of runs, because saturation slows the true fleet. \emph{Direct CP on
observed convergence times} meets R1 and R2, and every tolerance needs a new
calibration that costs $80$ times more simulation. \emph{Conformal
robustness of learned Lyapunov and barrier certificates}
\cite{hsu2025,lindemann2024,chee2023} certifies one fixed configuration of one
system, and a fleet needs a time-to-agreement.

\emph{Our idea} (Fig.~\ref{fig:concept}) is to let each tool supply the half
that the other leaves open. The \emph{annealed} Koopman operator, which averages the dynamics over
the random switching, is linear for any agent nonlinearity, and its spectrum
gives the decay rate. CP then calibrates the single constant from $200$ runs
of $50$ rounds. The calibration score is measured against the
\emph{estimated} rate, so an estimation error lengthens the deadline and
leaves the guarantee intact. To our knowledge, this is the first deadline
certificate for nonlinear consensus on randomly switching networks that
holds with a stated probability from finite data. The primary contributions
are:
\begin{itemize}\setlength{\itemsep}{1pt}
\item the \emph{conformal Koopman deadline} (Theorem~\ref{thm:deadline},
Corollary~\ref{cor:all_eps}), which meets R1 to R3 for any switching
distribution and any accuracy of the estimated operator;
\item the \emph{conformal decision certificate} (Theorem~\ref{thm:radius}), a
certified margin of the collective decision against corruption of the
transmitted data, derived from the eigenfunction of the same operator at
eigenvalue one; it corrects a nominal margin that worst-case perturbations
exceed in $13.3\%$ of test cases;
\item an evaluation against the four methods above on one metric per
requirement: coverage (R1), deadline length relative to the slowest run
(R2) and calibration cost (R3);
\item a test that detects spurious near-unit eigenvalues, which dictionary
enrichment creates and which lengthen the deadline $29$-fold.
\end{itemize}

\section{Preliminaries and Problem Formulation}
\subsection{Notation}
$\mathbb{N}=\{0,1,2,\dots\}$; $\mathbb{R}_{\ge0}$ and $\mathbb{R}_{>0}$ are
the reals $\ge0$ and $>0$; $\|\cdot\|$ is the Euclidean norm of a vector and
the matrix norm it induces, $\|\cdot\|_F$ the Frobenius norm,
$\|\cdot\|_\infty$ the supremum norm of a vector or a function and the
operator norm it induces; $\kappa(W)=\|W\|\|W^{-1}\|$ is the condition number
of an invertible matrix $W$; ${}^{\top}$ the transpose; $\mathbf{1}$ the
all-ones vector; $I$ the identity; $e_m$ the
$m$th basis vector; $\mathrm{diag}(v)$ the diagonal matrix with diagonal
$v$; $\otimes$ the Kronecker product; $\lceil\cdot\rceil$ the ceiling; $\log$
the natural logarithm; $\sgn$ the sign; $\nabla_y$ the gradient in $y$; $\Pr$
probability; $\E$ expectation; $\bar{x}=\tfrac1N\mathbf{1}^{\!\top}x$ the mean
of $x\in\mathbb{R}^N$. A hat marks a quantity estimated from data; the same
symbol alone is the true quantity. \emph{Acronyms:} EDMD (extended dynamic mode
decomposition), i.i.d.\ (independent and identically distributed). \emph{Baselines:} W (worst-case ergodicity), L (linearised spectral),
K (Koopman spectral with $C=1$), CP (direct conformal prediction); ours is CK
(conformal Koopman).

\subsection{Agent team on a time-evolving graph}
\begin{definition}[Markov-modulated agent team]\label{def:sys}
$N\in\mathbb{N}$ agents hold beliefs $x_i\in\mathbb{R}$, which form the team
belief $x=(x_1,\dots,x_N)\in\mathbb{R}^{N}$. A \emph{topology mode}
$m\in\mathcal{M}=\{1,\dots,M\}$ labels the communication graph active at the
current step, with weighted adjacency matrix
$A^{m}\in\mathbb{R}^{N\times N}_{\ge0}$, zero diagonal and directed Laplacian
$L_m=\mathrm{diag}(A^m\mathbf{1})-A^m$. Its entry $a^{m}_{ij}$ is the
influence of agent $j$ on agent $i$. With step size $h\in\mathbb{R}_{>0}$,
one communication round in mode $m$ maps belief $x_i$ to
\begin{equation}\label{eq:dyn}
  x_i^{+} = x_i + h\sum_{j=1}^{N} a^{m}_{ij}\,\tanh\!\left(x_j-x_i\right),
  \qquad i=1,\dots,N.
\end{equation}
We write $x^{+}=F_m(x)$. At step $k\in\mathbb{N}$ the belief is $x_k$ and
the mode $m_k$. The modes $(m_k)_{k\in\mathbb{N}}$ form an irreducible
aperiodic Markov chain, independent of $x$, with transition matrix
$P\in\mathbb{R}^{M\times M}$, $P_{mm'}=\Pr\{m_{k+1}=m'\mid m_k=m\}$, and
stationary distribution $\pi\in\mathbb{R}^M$, $\pi^{\!\top}P=\pi^{\!\top}$.
The joint state is
$z_k=(x_k,m_k)\in\mathcal{Z}=\mathbb{R}^{N}\times\mathcal{M}$.
\end{definition}

Each agent moves toward its current neighbours, which change at random; this
random sequence of graphs is the \emph{switching process}. The $\tanh$
coupling makes the pull of a neighbour saturate as the belief difference
grows, so \eqref{eq:dyn} is linear only near consensus. Progress is measured
by the disagreement $d_k := \|x_k-\bar{x}_k\mathbf{1}\|$, the distance of the
team belief from its mean $\bar{x}_k$. For a tolerance $\varepsilon\in(0,1)$
the convergence time
\begin{equation}\label{eq:T}
  T(\varepsilon) := \min\{k\in\mathbb{N} : d_k \le \varepsilon d_0\}
\end{equation}
is the first round at which the initial disagreement has shrunk by the
factor $\varepsilon$. We write $x_\infty\mathbf{1}:=\lim_{k\to\infty} x_k$, with
$x_\infty$ the common value of all beliefs. The mode keeps switching, so for
an observable $\varphi$ we write $\varphi(z_\infty):=\lim_{k\to\infty}\varphi(z_k)$
whenever this limit exists.

\begin{remark}\label{rem:random}
In general the Laplacians $L_m$ have different left null vectors, so
$x_\infty$ is a random variable that depends on the switching sequence as
well as on $x_0$. Statements about the collective decision therefore concern
$\E[x_\infty\mid z_0]$.
\end{remark}

\subsection{The annealed Koopman operator}
\begin{definition}[Annealed Koopman operator]\label{def:koop}
For bounded observables $g:\mathcal{Z}\to\mathbb{R}$, that is, bounded
functions of the joint state, define
\begin{align}\label{eq:koop}
  (\mathcal{K}g)(x,m) &:= \E\big[g(z_{k+1})\,\big|\,z_k=(x,m)\big] \nonumber\\
  &= \!\!\sum_{m'\in\mathcal{M}}\!\! P_{mm'}\, g\big(F_m(x),m'\big).
\end{align}
\end{definition}

\begin{lemma}\label{lem:markov}
$\mathcal{K}$ is linear, positive, satisfies $\|\mathcal{K}\|_\infty=1$ and
$\mathcal{K}\mathbf{1}_{\mathcal{Z}}=\mathbf{1}_{\mathcal{Z}}$ for the
constant function $\mathbf{1}_{\mathcal{Z}}\equiv1$ on $\mathcal{Z}$, and its
eigenvalues lie in the closed unit disc.
\end{lemma}
\begin{proof}
As a conditional expectation, \eqref{eq:koop} is linear and positive in $g$
and maps constants to themselves, so
$\mathcal{K}\mathbf{1}_{\mathcal{Z}}=\mathbf{1}_{\mathcal{Z}}$ and
$\|\mathcal{K}\|_\infty\ge1$. Positivity gives
$|\mathcal{K}g|\le\mathcal{K}|g|\le\|g\|_\infty$, so
$\|\mathcal{K}\|_\infty\le1$, and the spectral radius of a contraction is at
most one.
\end{proof}
The team is nonlinear, and $\mathcal{K}$ is a linear contraction. Its
subdominant eigenvalue $\lambda_2$, of largest modulus strictly inside the
unit disc, sets the rate $|\lambda_2|$ at which the fleet forgets an initial
disagreement.

\subsection{EDMD}
We estimate $\mathcal{K}$ by EDMD \cite{williams2015,korda2018}. Let
$V\in\mathbb{R}^{N\times q}$ hold the eigenvectors of the $\pi$-averaged
Laplacian $\sum_m\pi_mL_m$ for its $q$ eigenvalues of smallest modulus. The
slow collective coordinates $y=V^{\!\top}x\in\mathbb{R}^q$ form a
diffusion-map embedding \cite{coifman2006}. Let $B(x)$ be the row vector of
the $b$ monomials in $y$ up to degree $d$. The mode-indexed dictionary
$\Psi(x,m)=B(x)\otimes e_m^{\!\top}$ is a row vector of $bM$ observables; we
call $bM$ the \emph{lift dimension}, the size of the linear model. Stack the
rows $\Psi(z)$ at all observed states into a data matrix, also written
$\Psi$, and at their successor states into $\Psi'$. The estimate
$\widehat{K}=\arg\min_{K\in\mathbb{R}^{bM\times bM}}\|\Psi'-\Psi K\|_F$,
minimised over all $bM\times bM$ matrices $K$, is the matrix of
$\mathcal{K}$ on the span of the dictionary: it maps the coefficient vector
$c\in\mathbb{R}^{bM}$ of an observable $g(z)=\Psi(z)c$ to that of
$(\mathcal{K}g)(z)\approx\Psi(z)\widehat{K}c$. Its subdominant eigenvalue
$\widehat{\lambda}_2$ gives the estimated decay rate
$\widehat{\rho}:=|\widehat{\lambda}_2|$.

\begin{remark}\label{rem:lin}
Linearise $F_m(x)\approx\Phi_mx$ with $\Phi_m=I-hL_m$. On linear observables
$g(x,m)=c_m^{\!\top}x$, $c_m\in\mathbb{R}^N$, the annealed operator acts as
the $MN\times MN$ matrix
$c_m\mapsto\Phi_m^{\!\top}\sum_{m'}P_{mm'}c_{m'}$, whose subdominant modulus
$\rho_{\mathrm{L}}$ we compute in closed form; it defines baseline~L.
\end{remark}

\subsection{Split conformal prediction}\label{sec:cp}
Let $\mathcal{S}_{\mathrm{cal}}=\{\zeta_i\}_{i=1}^{n}$ be $n$ calibration
samples, here trajectories, and $\zeta_{\mathrm{new}}$ a new sample. A
\emph{nonconformity score} $s:\zeta\mapsto\mathbb{R}$ grows as a sample
departs from a model. For a miscoverage level $\delta\in(0,1)$ the conformal
quantile is
\begin{equation}\label{eq:quantile}
  q_{1-\delta} := s_{(\lceil(1-\delta)(n+1)\rceil)},
\end{equation}
where $s_{(1)}\le\dots\le s_{(n)}$ are the calibration scores
$s(\zeta_1),\dots,s(\zeta_n)$ sorted in increasing order and
$s_{(j)}:=+\infty$ for $j>n$. It is the $(1-\delta)$-quantile of the $n$
calibration scores together with the score of $\zeta_{\mathrm{new}}$, not yet
observed and counted as $+\infty$; for $n=200$ and $\delta=0.1$ it is the
$181$st smallest score.

\begin{lemma}[Split CP \cite{vovk2005,barber2023}]\label{lem:cp}
If $s(\zeta_1),\dots,s(\zeta_n),s(\zeta_{\mathrm{new}})$ are exchangeable,
that is, their joint distribution is invariant under permutations, then
$\Pr\{s(\zeta_{\mathrm{new}})\le q_{1-\delta}\}\ge1-\delta$, for any
dynamics.
\end{lemma}

\subsection{Problem formulation}
\begin{problem}\label{prob}
Given trajectory data from \eqref{eq:dyn}, with $A^m$, $P$, $h$ accessible
only through data, construct: (i) a deadline $\widehat{T}(\varepsilon)$
with $\Pr\{T(\varepsilon)\le\widehat{T}(\varepsilon)\}\ge1-\delta$ for a new
trajectory (R1), for any switching distribution and any accuracy of the
estimated model, that stays close to the largest convergence time (R2) and
holds for all $\varepsilon$ at once from one calibration (R3); (ii) a radius
$\widehat{\rho}_{\mathrm{CP}}$ such that, if the slow coordinates $y$ sent at
time zero are corrupted by a perturbation $\Delta y\in\mathbb{R}^q$ with
$\|\Delta y\|\le\widehat{\rho}_{\mathrm{CP}}$, the sign of
$\E[x_\infty\mid z_0]$, the binary decision the team will reach, stays the
same with probability at least $1-\delta$.
\end{problem}

\section{Conformal Koopman Certificates}
\subsection{The constant \texorpdfstring{$C$}{C}}
\begin{proposition}\label{prop:decay}
Let $\mathcal{G}$ be a $\mathcal{K}$-invariant subspace of observables vanishing
on the consensus manifold $\{x=\bar{x}\mathbf{1}\}$, with $\lambda_2$ the
spectral radius of the restriction $\mathcal{K}|_{\mathcal{G}}$. Then there is
$C\in\mathbb{R}_{>0}$ with
$\|\mathcal{K}^{k}|_{\mathcal{G}}\|\le C|\lambda_2|^{k}$ for all
$k\in\mathbb{N}$, where $\mathcal{K}^k$ is the $k$-fold composition of
$\mathcal{K}$, $(\mathcal{K}^kg)(z)=\E[g(z_k)\mid z_0=z]$.
For a diagonalisable finite-dimensional compression with eigenvector matrix
$W$ one may take $C=\kappa(W)$.
\end{proposition}
\begin{proof}
With the compression $W\Lambda W^{-1}$, $\Lambda$ the diagonal matrix of its
eigenvalues,
$\|W\Lambda^kW^{-1}\|\le\|W\|\,\|\Lambda^k\|\,\|W^{-1}\|=\kappa(W)|\lambda_2|^k$
on $\mathcal{G}$.
\end{proof}

Proposition~\ref{prop:decay} is the classical route. On our instances
$\kappa(W)$ ranged from $56.8$ to $190.8$, which inflates the deadline to
$1.87$ to $3.24$ times the observed worst case. Setting $C=1$ is common; it
replaces the theorem by an assumption. We calibrate $C$.

\subsection{Conformal Koopman deadline}
Fix a burn-in $k_0\in\mathbb{N}$, the number of initial rounds treated as
transient, and a calibration horizon $K_{\mathrm{cal}}>k_0$. For a trajectory
$\zeta$ we instantiate the nonconformity score $s$ of Section~\ref{sec:cp} as
\begin{equation}\label{eq:score}
  s(\zeta) := \max_{k_0\le k\le K_{\mathrm{cal}}}
  \frac{d_k}{\widehat{\rho}^{\,k}d_0},
\end{equation}
the smallest constant $C$ for which the estimated envelope
$d_k\le C\widehat{\rho}^{\,k}d_0$ holds along $\zeta$ for
$k_0\le k\le K_{\mathrm{cal}}$.

\begin{assumption}[Exchangeability]\label{as:exch}
The scores of the calibration trajectories and of a new trajectory are
exchangeable. This holds when initial beliefs and initial modes are drawn
i.i.d.\ from one fixed distribution, for example the one used at
commissioning.
\end{assumption}

\begin{assumption}[Post-transient dominance]\label{as:tail}
With probability one,
$d_{k+1}/(\widehat{\rho}^{\,k+1}d_0)\le d_k/(\widehat{\rho}^{\,k}d_0)$ for
all $k\ge k_0$.
\end{assumption}

Assumption~\ref{as:tail} lets a calibration on $K_{\mathrm{cal}}$-round
trajectories cover tolerances reached many rounds later. It reflects that the
sample-path decay rate of a randomly switching system is in general strictly
smaller than the annealed rate $|\lambda_2|$; Section~\ref{sec:num} checks it
empirically.

\begin{theorem}[Conformal Koopman deadline]\label{thm:deadline}
Let $q_{1-\delta}$ be the conformal quantile \eqref{eq:quantile} of the scores
\eqref{eq:score} over $n$ calibration trajectories. Define
\begin{equation}\label{eq:TCK}
  \widehat{T}_{\mathrm{CK}}(\varepsilon) :=
  \left\lceil \frac{\log(\varepsilon/q_{1-\delta})}{\log\widehat{\rho}}\right\rceil .
\end{equation}
Under Assumption~\ref{as:exch}, a new trajectory satisfies
\begin{equation}\label{eq:env}
  \Pr\Big\{d_k\le q_{1-\delta}\,\widehat{\rho}^{\,k}d_0,\
  \forall k\in[k_0,K_{\mathrm{cal}}]\Big\}\ \ge\ 1-\delta .
\end{equation}
If Assumption~\ref{as:tail} also holds, then \eqref{eq:env} extends to all
$k\ge k_0$, and
$\Pr\{T(\varepsilon)\le\widehat{T}_{\mathrm{CK}}(\varepsilon)\}\ge1-\delta$
for every $\varepsilon$ with $\widehat{T}_{\mathrm{CK}}(\varepsilon)\ge k_0$.
\end{theorem}
\begin{proof}
The score \eqref{eq:score} is a measurable function of a trajectory, so
Assumption~\ref{as:exch} and Lemma~\ref{lem:cp} give
$\Pr\{s(\zeta_{\mathrm{new}})\le q_{1-\delta}\}\ge1-\delta$, which is the
event in \eqref{eq:env}. Under Assumption~\ref{as:tail} the maximum over
$[k_0,K_{\mathrm{cal}}]$ equals the supremum over $[k_0,\infty)$, so the
envelope extends. On that event,
$d_k\le q_{1-\delta}\widehat{\rho}^{\,k}d_0\le\varepsilon d_0$ as soon as
$\widehat{\rho}^{\,k}\le\varepsilon/q_{1-\delta}$, that is, by
\eqref{eq:TCK}, for $k\ge\widehat{T}_{\mathrm{CK}}(\varepsilon)$.
\end{proof}

\begin{corollary}[Simultaneity in $\varepsilon$]\label{cor:all_eps}
The conclusion of Theorem~\ref{thm:deadline} holds jointly over all
admissible $\varepsilon$, that is, all $\varepsilon$ with
$\widehat{T}_{\mathrm{CK}}(\varepsilon)\ge k_0$:
$\Pr\{\forall\varepsilon:\,T(\varepsilon)\le
\widehat{T}_{\mathrm{CK}}(\varepsilon)\}\ge1-\delta$.
\end{corollary}
\begin{proof}
All these statements follow from the single event \eqref{eq:env}.
\end{proof}

\begin{remark}[Estimation error is absorbed]\label{rem:absorb}
The coverage guarantee of Theorem~\ref{thm:deadline} holds for every value
of $\widehat{\rho}$, at any distance from $|\lambda_2|$. The same
$\widehat{\rho}$ appears in \eqref{eq:score} and \eqref{eq:TCK}, so any error
in $\widehat{\rho}$ is absorbed by $q_{1-\delta}$: it lengthens the deadline
and leaves R1 intact, independently of eigenvalue perturbation bounds
\cite{bauer1960}.
\end{remark}

\subsection{Conformal decision certificate}
\begin{lemma}\label{lem:mart}
If $\varphi_1$ is an eigenfunction with $\mathcal{K}\varphi_1=\varphi_1$ that
varies with $x$, then $\varphi_1(z_k)$ is a bounded martingale and
$\varphi_1(z_0)=\E[\varphi_1(z_\infty)\mid z_0]$. Since $x_k$ approaches the
consensus manifold, $\varphi_1$ equals $\E[x_\infty\mid z_0]$ up to an affine
normalisation.
\end{lemma}
\begin{proof}
$\E[\varphi_1(z_{k+1})\mid z_k]=(\mathcal{K}\varphi_1)(z_k)=\varphi_1(z_k)$
is the martingale property. Boundedness and martingale convergence give the
first claim, and restriction to the consensus manifold the second.
\end{proof}

Physically, $\varphi_1$ is the weighted opinion that the switching dynamics
conserve in expectation: the answer the team is committed to at time zero.
The binary decision is $D(z_0)=\sgn\varphi_1(z_0)$. For perturbations
$\Delta y\in\mathbb{R}^q$ of the transmitted coordinates $y=V^{\!\top}x_0$,
write $z_0+\Delta y$ for the joint state with initial belief $x_0+V\Delta y$.
The nominal first-order radius
\begin{equation}\label{eq:radius}
  \widehat{\rho}(z_0) := \frac{|\widehat{\varphi}_1(z_0)|}
  {\|\nabla_{y}\widehat{\varphi}_1(z_0)\|}
\end{equation}
is the distance of the estimated functional from its decision boundary in
units of its slope; the argument $z_0$ distinguishes it from the decay rate
$\widehat{\rho}$.

\begin{theorem}[Conformal robustness radius]\label{thm:radius}
Let $\varphi^{\star}(z_0):=\E[x_\infty\mid z_0]$ be the true decision
functional, with $\widehat{\varphi}_1$ affinely normalised to it. Let
$z_0^{(1)},\dots,z_0^{(n)}$ be calibration initial states with scores
$s_i=|\varphi^{\star}(z_0^{(i)})-\widehat{\varphi}_1(z_0^{(i)})|/
\|\nabla_y\widehat{\varphi}_1(z_0^{(i)})\|$ and conformal quantile
$q_{1-\delta}$. Set
$\widehat{\rho}_{\mathrm{CP}}:=\max\{\widehat{\rho}(z_0)-q_{1-\delta},0\}$.
Then, to first order in $\|\Delta y\|$ and under Assumption~\ref{as:exch},
\begin{equation}\label{eq:radius_guarantee}
  \Pr\big\{\sgn\varphi^{\star}(z_0+\Delta y)=D(z_0),\
  \forall\,\|\Delta y\|\le\widehat{\rho}_{\mathrm{CP}}\big\}\ge1-\delta .
\end{equation}
\end{theorem}
\begin{proof}
The score $s_i$ is the estimation error of $\widehat{\varphi}_1$ at
$z_0^{(i)}$ in units of its slope. By Lemma~\ref{lem:cp}, with probability at
least $1-\delta$,
$|\varphi^{\star}(z_0)-\widehat{\varphi}_1(z_0)|\le
q_{1-\delta}\|\nabla_y\widehat{\varphi}_1(z_0)\|$. The reverse triangle
inequality and \eqref{eq:radius} give
$|\varphi^{\star}(z_0)|\ge
\widehat{\rho}_{\mathrm{CP}}\|\nabla_y\widehat{\varphi}_1(z_0)\|$, so
$\sgn\varphi^{\star}(z_0)=D(z_0)$ whenever $\widehat{\rho}_{\mathrm{CP}}>0$.
To first order,
$|\varphi^{\star}(z_0+\Delta y)-\varphi^{\star}(z_0)|\le
\|\nabla_y\widehat{\varphi}_1(z_0)\|\,\|\Delta y\|$, which is strictly smaller
than $|\varphi^{\star}(z_0)|$ for $\|\Delta y\|<\widehat{\rho}_{\mathrm{CP}}$,
so the sign is kept. This is \eqref{eq:radius_guarantee}.
\end{proof}

\section{Numerical Results}\label{sec:num}
\renewcommand{\thesubsubsection}{\thesection-\arabic{subsubsection}}
We simulate a fleet of $N=60$ agents that switches at random among $M=3$
communication graphs, keeping the current one with probability
$P_{mm}=0.7$, with step size $h=0.02$. One round is one exchange of messages; at $20$
rounds per second, $100$ rounds take $5$\,s. Initial beliefs are i.i.d.\
$\mathcal{N}(0,1)$, so two agents start a mean distance
$2/\sqrt{\pi}\approx1.13$ apart, where $\tanh$ pulls $28\%$ less than its
linearisation; the initial mode is uniform on $\mathcal{M}$. A
\emph{network} is one draw of the $M$ graphs, each linking every ordered
pair of agents independently with probability $p_{\mathrm{e}}=0.10$, so that
each agent hears from about six others, with a weight uniform on
$[0.5,1.5]$. EDMD uses $q=5$ slow coordinates and degree $d=2$ (lift
$bM=63$) on $400$ runs of $120$ steps; CK uses $\delta=0.1$, $k_0=20$,
$K_{\mathrm{cal}}=50$ and $n=200$ calibration runs. We ran every experiment on three independent
networks, so that no conclusion rests on one draw; Table~\ref{tab:cov} and
Fig.~\ref{fig:deadline} show network~1. On network~1 the fitted decay rate
$\widehat{\rho}=0.961$ ($0.965$ and $0.958$ on networks~2 and~3) removes
about $4\%$ of the remaining disagreement per round, so the disagreement
halves every $17.5$ rounds, or $0.9$\,s. The calibrated constant
$q_{1-\delta}=0.615$ ($0.629$, $0.696$) means that in $90\%$ of calibration
runs the disagreement stayed below $62\%$ of the Koopman
prediction with $C=1$ from round $20$ to $50$.

We compare our deadline (CK) with four alternatives. The worst-case
bound~W assumes that each agent always listens to its neighbours as little as
the model allows. For belief differences up to $D_0=6$, six standard
deviations, the $\tanh$ coupling keeps at least the fraction
$\theta_{\min}=\tanh(D_0)/D_0=0.167$ of the linear pull, so with smallest
edge weight $a_{\min}=0.5$ every positive weight in the averaging
\eqref{eq:dyn} is at least $\eta=h\,a_{\min}\theta_{\min}=1.7\times10^{-3}$.
Every graph of our networks is rooted (one agent reaches all others along
directed edges), and any $N-1$ rooted graphs with self-loops compose to a
graph in which one agent reaches all others directly \cite{cao2008}. Every
block of $N-1=59$ rounds therefore shrinks the spread
$\max_ix_i-\min_ix_i$ by at least the factor $1-\eta^{59}$, with
$\eta^{59}\approx10^{-164}$, for every switching sequence, and the guaranteed
deadline $\widehat{T}_{\mathrm{W}}$, of order
$(N-1)\log(1/\varepsilon)/\eta^{N-1}$, exceeds $10^{164}$ rounds. L and K
evaluate \eqref{eq:TCK} with $C=1$ in place of $q_{1-\delta}$, using the rate
$\rho_{\mathrm{L}}$ of the linearised team and the fitted rate
$\widehat{\rho}$, giving $\widehat{T}_{\mathrm{L}}$ and
$\widehat{T}_{\mathrm{K}}$; CP takes the conformal quantile of the
convergence times of the calibration runs, giving
$\widehat{T}_{\mathrm{CP}}$.

\subsubsection{When the fleet agrees}
\begin{figure*}[t]
\centering
\includegraphics[width=0.92\textwidth]{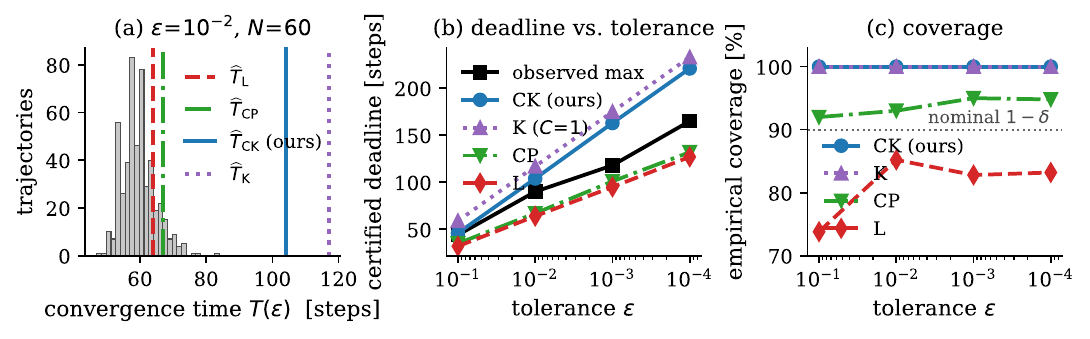}
\caption{(a) Convergence times at $\varepsilon=10^{-2}$ and the four
data-driven deadlines. (b) Certified deadline against tolerance; CK tracks the
observed maximum from one calibration. (c) Coverage against the nominal
$90\%$ level; L falls below it.}
\label{fig:deadline}
\end{figure*}

Table~\ref{tab:cov} and Fig.~\ref{fig:deadline} compare the deadlines on
$500$ test runs at tolerances from $\varepsilon=10^{-1}$ (disagreement
reduced tenfold) to $10^{-4}$ (ten-thousandfold). Observed max is the
convergence time of the slowest run; coverage (cov.) is the percentage of
runs that had agreed by the deadline, with target $90\%$. At
$\varepsilon=10^{-2}$ the slowest run agreed by round $90$. L allows action
at round $64$ ($3.2$\,s), and in $14.8\%$ of runs, about one mission in
seven, the fleet would act on a split opinion. CK allows action at round
$104$ ($5.2$\,s), $14$ rounds after the slowest run; these $2$\,s beyond L
buy the $90\%$ guarantee.

The deadline of L comes early because L assumes that the pull between two
agents grows in proportion to their disagreement. In the real fleet the pull
saturates, so early in a run, when beliefs differ widely, the fleet
converges more slowly than L predicts: L halves the disagreement every $9.5$
rounds ($\rho_{\mathrm{L}}=0.930$; $0.944$ and $0.935$ on networks~2 and~3),
the fitted rate every $17.5$. The effect depends on the network: across
tolerances L covered $73.8\%$ to $85.2\%$ of runs on network~1, $99.0\%$ to
$99.8\%$ on network~2 and $90.6\%$ to $95.8\%$ on network~3, and L itself
leaves open which case applies. K, the fitted rate with an assumed $C=1$,
met the target here, with a deadline $34\%$ above the slowest run at
$\varepsilon=10^{-1}$, against $5\%$ for CK. CP meets the target ($90.2\%$
to $95.2\%$) with the shortest deadlines among the methods that do, at a
cost: each calibration run must continue until the fleet reaches the target
tolerance, here $4000$ rounds ($3.3$\,min), so CP needs $8\times10^{5}$
simulated rounds per tolerance and a new calibration whenever the mission
lowers its tolerance. CK needs $10^{4}$ rounds once ($200$ runs of
$2.5$\,s) and reads every deadline from \eqref{eq:TCK}. CK met its deadline
in $99.6\%$ to $100\%$ of runs on all three networks. Its envelope was
crossed at some round in $7.0\%$ to $13.4\%$ of runs, close to the $10\%$
the calibration allows; these mid-run crossings delayed agreement past the
deadline in at most $0.4\%$ of runs. All observations were consistent with
Assumption~\ref{as:tail}.

\begin{table}[t]
\caption{Certified deadlines (rounds) and coverage (\%), network~1}
\label{tab:cov}
\centering\footnotesize
\setlength{\tabcolsep}{3.2pt}
\begin{tabular}{@{}lrrrrrrrrr@{}}
\toprule
& \multicolumn{2}{c}{$\varepsilon\!=\!10^{-1}$} &
  \multicolumn{2}{c}{$\varepsilon\!=\!10^{-2}$} &
  \multicolumn{2}{c}{$\varepsilon\!=\!10^{-3}$} &
  \multicolumn{2}{c}{$\varepsilon\!=\!10^{-4}$} &
  calibration \\
\cmidrule(lr){2-3}\cmidrule(lr){4-5}\cmidrule(lr){6-7}\cmidrule(lr){8-9}
method & $\widehat{T}$ & cov. & $\widehat{T}$ & cov. & $\widehat{T}$ & cov.
& $\widehat{T}$ & cov. & steps \\
\midrule
observed max & 44 &  & 90 &  & 118 &  & 165 &  &  \\
W  & \multicolumn{8}{c}{$>10^{164}$ at every $\varepsilon$ (deterministic)} & 0 \\
L  & 32 & 73.8 & 64 & 85.2 & 95 & 82.8 & 127 & 83.2 & 0 \\
K  & 59 & 100 & 117 & 100 & 175 & 100 & 233 & 100 & $10^{4}$ \\
CP & 35 & 92.0 & 67 & 93.0 & 101 & 95.0 & 132 & 94.8 & $8{\cdot}10^{5}$ \\
\textbf{CK} & \textbf{46} & \textbf{100} & \textbf{104} & \textbf{100}
& \textbf{163} & \textbf{100} & \textbf{221} & \textbf{100} & $\mathbf{10^{4}}$ \\
\bottomrule
\end{tabular}
\end{table}

\subsubsection{Larger fleets}
Table~\ref{tab:scale} and Fig.~\ref{fig:scale}(a) repeat the test for
$N=20$ to $240$ agents, with $150$ test runs each at $\varepsilon=10^{-2}$
and about six neighbours per agent (edge probability $6/N$). $T_{\max}$ is
the slowest run; violation of K and coverage of L are the percentages of runs
that missed $\widehat{T}_{\mathrm{K}}$ and that met
$\widehat{T}_{\mathrm{L}}$. The linear model keeps the same $63$ basis functions
at every size, because the lift dimension $bM$ depends only on $q$ and $d$; fitting
takes $144$ to $162$\,ms and calibration the same $10^{4}$ rounds, so
certifying $240$ robots costs as much as certifying $20$. The deadline grows from $94$
rounds at $N=20$ to $164$ at $N=240$ (peak $170$ at $N=160$): with six
neighbours per agent, a message needs more hops to cross a larger fleet.
Coverage of L rises from $69.3\%$ to $100\%$: in a larger fleet each agent
hears neighbours on both sides of its own belief, their pulls partly cancel,
the differences stay small, and saturation fades. The same linearised
computation thus gives a deadline that every run meets at $N=240$ and
$30.7\%$ of runs miss at $N=20$, and L itself leaves open which case
applies. CK takes its rate and constant from runs of the actual, saturating
fleet, so Theorem~\ref{thm:deadline} holds in both cases, where saturation
matters ($N=20$) and where it fades ($N=240$).

\begin{table}[t]
\caption{Scaling in team size at $\varepsilon=10^{-2}$}
\label{tab:scale}
\centering\footnotesize
\setlength{\tabcolsep}{3.5pt}
\begin{tabular}{@{}rccrrrrrr@{}}
\toprule
$N$ & lift & $\widehat{\rho}$ & $\widehat{T}_{\mathrm{K}}$ &
$\widehat{T}_{\mathrm{L}}$ & $T_{\max}$ &
\begin{tabular}[b]{@{}c@{}}violation\\of K (\%)\end{tabular} &
\begin{tabular}[b]{@{}c@{}}coverage\\of L (\%)\end{tabular} &
\begin{tabular}[b]{@{}c@{}}fit\\(ms)\end{tabular} \\
\midrule
20  & 63 & 0.9520 & 94  & 54  & 69  & 0.0 & 69.3  & 144 \\
60  & 63 & 0.9579 & 108 & 64  & 76  & 0.0 & 88.0  & 148 \\
100 & 63 & 0.9719 & 162 & 85  & 89  & 0.0 & 97.3  & 144 \\
160 & 63 & 0.9732 & 170 & 104 & 106 & 0.0 & 99.3  & 159 \\
240 & 63 & 0.9722 & 164 & 101 & 98  & 0.0 & 100.0 & 162 \\
\bottomrule
\end{tabular}
\end{table}

\subsubsection{How much corruption the decision absorbs}\label{sec:numdec}
The fleet's decision is the sign of its predicted final value: whether it
will settle above or below zero. The predictor $\widehat{\varphi}_1$ comes
from the linear model. Eigenvalue one of $\widehat{K}$ also belongs to the
constant function (Lemma~\ref{lem:markov}), which predicts the same outcome
for every start, and Lemma~\ref{lem:mart} needs an eigenfunction that varies
with $x$. Each eigenvector $c$ of $\widehat{K}$ gives an eigenfunction
$\Psi(z)c$ with constant, linear and quadratic parts; since for the
linearised team the outcome $\E[x_\infty\mid z_0]$ is a weighted average of
the initial beliefs, we took the one with eigenvalue nearest one among those
dominated by the linear part (eigenvalue $1.0003$ compressed, $1.0002$
full). We scaled it against Monte-Carlo averages of the final value over $50$
switching sequences per start, calibrated Theorem~\ref{thm:radius} on $40$
starts and tested on $30$ others, corrupting the transmitted data by exactly
the margin in the most harmful direction,
$-\sgn(\widehat{\varphi}_1)\nabla_y\widehat{\varphi}_1$, which moves the
prediction fastest toward zero.

\begin{figure}[t]
\centering
\includegraphics[width=\columnwidth]{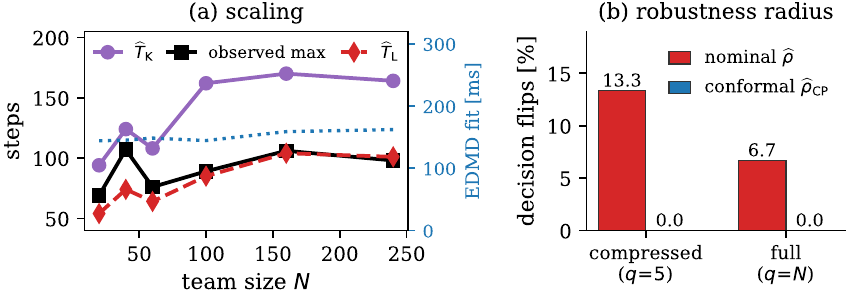}
\caption{(a) Deadlines and EDMD fit time against team size. (b) Decision
flips under worst-case perturbation at the nominal radius \eqref{eq:radius}
and at the conformal radius of Theorem~\ref{thm:radius}, which flipped no
decision in either case.}
\label{fig:scale}
\end{figure}

With five transmitted coordinates ($q=5$, lift $63$), the nominal margin
\eqref{eq:radius} flipped the decision in $13.3\%$ of test cases. The certified margin averaged $0.482$ against a nominal $1.108$
($43.5\%$), in belief units, and flipped zero decisions
(Fig.~\ref{fig:scale}(b)). With all $60$ coordinates ($q=N$, lift $183$) the
nominal margin flipped $6.7\%$; the certified one, $0.525$ against $0.778$
($67.5\%$), flipped zero. Two facts limit the margin. In $6.7\%$ of cases the
predicted decision already differed from the true one before any
corruption, the residual that Theorem~\ref{thm:radius} allows. And the error
of $\widehat{\varphi}_1$ comes from the coordinates themselves: tripling the
data, or enlarging the dictionary from $q=5$ to $q=12$, kept it between
$28\%$ and $42\%$, against a Monte-Carlo error of $0.6\%$. Physically, which
agents dominate the outcome depends on graph details that five collective
coordinates average out (Remark~\ref{rem:random}).

\subsubsection{Larger dictionaries and spurious slow patterns}
A larger dictionary can mislead the fit. With polynomials of degree $1$ and
$2$ the deadline at $\varepsilon=10^{-2}$ was $107$ and $117$ rounds. With
degree $3$ (lift $138$) the fit reported an eigenvalue at $0.99864$, a
pattern that would take about $740$ rounds ($37$\,s) to fade, and the
deadline jumped to $3393$ rounds. The real fleet settles within $100$ rounds
($5$\,s), so this pattern is an artefact that degree-3 terms mimic on the
finite data. The warning sign is the condition number $\kappa(W)$ of the
eigenvector matrix, which measures how much errors in $\widehat{K}$ move the
computed eigenvalues: it rose sevenfold, from $168$ to $1168$. We recommend
the largest dictionary before $\kappa(W)$ jumps; residual-based spectral
verification \cite{colbrook2024} offers a remedy with guarantees. By
Remark~\ref{rem:absorb} such an artefact lengthens the deadline and leaves
R1 intact.

\section{Conclusion}
From one calibration on $50$-round trajectories, conformal Koopman
certification gives deadlines for every tolerance and decision margins, at $1/80$ of the
calibration cost of direct conformal prediction, independent of team size. Three questions remain open: the guarantee is marginal,
that is, it holds on average over initial conditions;
Assumption~\ref{as:tail} awaits a proof; and mobile teams induce
state-dependent switching, which breaks the product structure of
\eqref{eq:koop}.


\begin{thebibliography}{99}
\bibitem{koopman1931} B.~O. Koopman, ``Hamiltonian systems and transformation in Hilbert space,'' \emph{Proc. Nat. Acad. Sci.}, vol.~17, no.~5, pp.~315--318, 1931.
\bibitem{mezic2005} I.~Mezi\'c, ``Spectral properties of dynamical systems, model reduction and decompositions,'' \emph{Nonlinear Dyn.}, vol.~41, pp.~309--325, 2005.
\bibitem{williams2015} M.~O. Williams, I.~G. Kevrekidis, and C.~W. Rowley, ``A data-driven approximation of the Koopman operator: extending dynamic mode decomposition,'' \emph{J. Nonlinear Sci.}, vol.~25, pp.~1307--1346, 2015.
\bibitem{korda2018} M.~Korda and I.~Mezi\'c, ``On convergence of extended dynamic mode decomposition to the Koopman operator,'' \emph{J. Nonlinear Sci.}, vol.~28, pp.~687--710, 2018.
\bibitem{colbrook2024} M.~J. Colbrook and A.~Townsend, ``Rigorous data-driven computation of spectral properties of Koopman operators for dynamical systems,'' \emph{Commun. Pure Appl. Math.}, vol.~77, no.~1, pp.~221--283, 2024.
\bibitem{olfati2004} R.~Olfati-Saber and R.~M. Murray, ``Consensus problems in networks of agents with switching topology and time-delays,'' \emph{IEEE Trans. Autom. Control}, vol.~49, no.~9, pp.~1520--1533, 2004.
\bibitem{jadbabaie2003} A.~Jadbabaie, J.~Lin, and A.~S. Morse, ``Coordination of groups of mobile autonomous agents using nearest neighbor rules,'' \emph{IEEE Trans. Autom. Control}, vol.~48, no.~6, pp.~988--1001, 2003.
\bibitem{ren2005} W.~Ren and R.~W. Beard, ``Consensus seeking in multiagent systems under dynamically changing interaction topologies,'' \emph{IEEE Trans. Autom. Control}, vol.~50, no.~5, pp.~655--661, 2005.
\bibitem{nedic2015} A.~Nedi\'c and A.~Olshevsky, ``Distributed optimization over time-varying directed graphs,'' \emph{IEEE Trans. Autom. Control}, vol.~60, no.~3, pp.~601--615, 2015.
\bibitem{cao2008} M.~Cao, A.~S. Morse, and B.~D.~O. Anderson, ``Reaching a consensus in a dynamically changing environment: a graphical approach,'' \emph{SIAM J. Control Optim.}, vol.~47, no.~2, pp.~575--600, 2008.
\bibitem{coifman2006} R.~R. Coifman and S.~Lafon, ``Diffusion maps,'' \emph{Appl. Comput. Harmon. Anal.}, vol.~21, no.~1, pp.~5--30, 2006.
\bibitem{vovk2005} V.~Vovk, A.~Gammerman, and G.~Shafer, \emph{Algorithmic Learning in a Random World}. New York, NY, USA: Springer, 2005.
\bibitem{angelopoulos2021} A.~N. Angelopoulos and S.~Bates, ``A gentle introduction to conformal prediction and distribution-free uncertainty quantification,'' 2021, \emph{arXiv:2107.07511}.
\bibitem{barber2023} R.~F. Barber, E.~J. Cand\`es, A.~Ramdas, and R.~J. Tibshirani, ``Conformal prediction beyond exchangeability,'' \emph{Ann. Statist.}, vol.~51, no.~2, pp.~816--845, 2023.
\bibitem{hsu2025} T.-W. Hsu and H.~Tsukamoto, ``Statistical guarantees in data-driven nonlinear control: conformal robustness for stability and safety,'' \emph{IEEE Control Syst. Lett.}, vol.~9, pp.~997--1002, 2025.
\bibitem{lindemann2024} L.~Lindemann, Y.~Zhao, X.~Yu, G.~J. Pappas, and J.~V. Deshmukh, ``Formal verification and control with conformal prediction,'' 2024, \emph{arXiv:2409.00536}.
\bibitem{chee2023} K.~Y. Chee, M.~A. Hsieh, and G.~J. Pappas, ``Uncertainty quantification for learning-based MPC using weighted conformal prediction,'' in \emph{Proc. IEEE CDC}, 2023, pp.~342--349.
\bibitem{bauer1960} F.~L. Bauer and C.~T. Fike, ``Norms and exclusion theorems,'' \emph{Numer. Math.}, vol.~2, pp.~137--141, 1960.
\end{thebibliography}
\end{document}